# HIERARCHICAL LOCATION SERVICE WITH PREDICTION IN MOBILE AD-HOC NETWORKS *

Ebtisam Amar[1], Selma Boumerdassi[2] and Éric Renault[3,4]

[1] University of Pierre and Marie Curie, Paris, France
ebtisam.amar@etu.upmc.fr
[2] CNAM – CEDRIC, Paris, France
selma.boumerdassi@cnam.fr
[3] Institut Télécom – Télécom SudParis, Évry, France
[4] Samovar UMR INT-CNRS 5157, Évry, France
eric.renault@it-sudparis.eu

## ABSTRACT

Position-based routing protocols take advantage of location information to perform a stateless and efficient routing. To enable position-based routing, a node must be able to discover the location of the messages' destination node. This task is typically accomplished by a location service. Recently, several location service protocols have been developed for ad hoc networks.

In this paper we propose a novel location service called PHLS: Predictive Hierarchical Location Service. In PHLS, the entire network is partitioned into a hierarchy of smaller and smaller regions. For each node, one node in each-level region of the hierarchy is chosen as its local location server. When the network initializes or when a node attaches the network, nodes contact their local location server with their current location information (ie. position and velocity). Then, they only need to update their location server when they move away from their current region. Finally, nodes query their location servers and get the exact or predicted location of destination nodes.

## 1. INTRODUCTION

A mobile ad-hoc network (MANET) is a self-organizing and self-configuring multi-hop wireless network, where the network structure changes dynamically due to node mobility. In such a network, each node operates not only as a host, but also as a router, forwarding packets for other mobile nodes.

One of the biggest challenges in MANETs is the design of scalable routing protocols. As of today, existing routing protocols are either based on topology or location.

Topology-based routing protocols assume no knowledge of the mobile nodes' positions. Proactive protocols such as DSDV [1], reactive protocols like DSR [2] and AODV [3], and hybrid protocols such as ZRP [4] are typically based on topology. They rely on discovering and maintaining global states to route packets. As a result, their scalability is limited.

---

Position-based routing protocols [5, 6] have been proposed to address the scalability problem associated with early ad-hoc routing protocols. They use nodes' location information available from positioning systems such as Global Positioning System (GPS) to forward packets. Since local information (neighbour's locations) is used for global routing, position-based routing protocols can potentially scale to larger number of nodes than topology-based routing protocols. However, with a position-based routing protocol, the source node needs to retrieve the location of the destination node before sending a packet.

Therefore, one of the main challenges in position-based routing protocols has been the design of different location services to track the location of mobile nodes and reply to location queries efficiently. This paper aims at proposing a novel hierarchical location service which main enhancement over other solutions is the use of prediction information for mobile nodes.

The rest of the paper is organized as follows. The next section gives a brief description of existing location services. Then, Sec. 3 presents our location service and a qualitative comparison with two pre-existing location services is provided in Sec. 4. Section 5 provides a mathematical analysis of our proposal to demonstrate its scalability while Sec. 6 presents an evaluation of its performance using simulations. Finally, Sec. 7 concludes the paper and points out some future research issues.

## 2. RELATED WORKS

There is a growing body of work that uses location information for routing in MANETs and numerous protocols for location services [7, 8, 9, 10, 11, 12, 13, 14, 15, 16, 17, 18, 19] have been proposed to solve location tracking. [20] provides a recent survey and presents a qualitative comparison of different location services found in the literature of MANETs, while [21, 22] provide a realistic, quantitative analysis comparing the performance of some location services.

Location services can be divided into several categories: hierarchal or flat depending on whether a hierarchy of recursively defined subareas are used, with or without prediction, etc. The following focuses on the presentation of HLS and PLS, two location services that are of great help to introduce our work in the next section.

### 2.1. Hierarchical Location Service

Hierarchical Location Service (HLS) [14] partitions the area containing the ad-hoc network into cells and regions, this partition being known by all participating nodes.

Cells are grouping nodes at the lowest level, and it is assumed that all nodes in a given cell can directly communicate with all the other nodes in the same cell. Then, cells are grouped into regions which are grouped into larger regions, thus creating the hierarchy. Regions composed of cells are level-1 regions, regions composed of level-1 regions are level-2 regions and so on. Regions at the same level do not intersect, ie. each cell is part of one and only one level-1 region and each level-n region is a member of exactly one level-$(n + 1)$ region.

For a given node, one specific cell is selected for each level of the hierarchy by means of a hash function. As the node changes its position, it transmits position updates to these responsible cells. If another node wants to find the position of the node, it uses the same hash function to determine the cells that may hold the position information of the node. It then queries the nodes in these cells in the hierarchy order until it receives a reply containing the current position of the node.

## 2.2. Predictive Location Service

Predictive Location Service (PLS) [15] is a flat location service that acts according to the following two design guidelines: i) prediction, ie. saved information is used to estimate the mobile node's current location and ii) adaptability, ie. several parameters adapt to the mobile node's speed.

In PLS, location information are periodically shared among the neighbors and a mobile node floods the network if the required location information is not found in the mobile node's location table. A mobile node's previous state (both location and velocity) is used to predict the mobile node's current state.

Regularly, a mobile node sends a Location Packet (LP) to all of its 1-hop-neighbors, ie. the mobile node transmits a broadcast packet with time-to-live (TTL) set to one. Each LP contains the current location information (both coordinates and velocity of the node) of the source mobile node plus location information about other mobile nodes that are stored in the source mobile node's location table. Every time a mobile node receives an LP, it updates its location table by comparing its own table to the location information in the LP.

Location entries that are older than a specific amount of time are considered as too old and are deleted from the location table.

When a location query is received by a mobile node and there is an entry for the searched mobile node in the queried mobile node's location table, the queried mobile node predicts the location of the searched mobile node using the previously saved location and velocity of the searched mobile node.

# 3. PREDICTIVE-HIERARCHICAL LOCATION SERVICE

Our novel algorithm, Predictive Hierarchal Location Service (PHLS), is based on a hierarchical location service protocol as such protocols scale better than flat location service protocols and we expect to enhance the location request answer by introducing prediction. In this way, we propose to combine hierarchy to achieve scalability with prediction to enhance the performance.

As presented above for HLS, PHLS divides the communication area into cells that are grouped into regions, and regions into larger regions, etc. Moreover, the behaviour of nodes in PHLS when the network initializes or when nodes attach the network remains the same.

In the following, we assume each node has a unique ID and is aware of its own position through the support of GPS or any other positioning technique.

## 3.1. Area Division

PHLS partitions must be known to all participating nodes. The shape and size of these regions can be chosen arbitrary according to the properties of the network. Nodes in a given cell must be able to send packets to all other nodes in the same cell. This can be achieved by choosing an appropriate cell size, where the distance between any two nodes in the cell must be smaller than the radio range. Cells and then regions are grouped hierarchically into regions of different levels, where the top-level forms the whole network.

### 3.2. Location Servers Selection

We assume that the location server density is one server in each level-n region in the network.

After the network is initialized and the partition is done, each node needs to designate some nodes, namely its location servers and update them with its location.

A hash function is used to map the node's ID to a set of location servers. All nodes select one location server in each level of the hierarchy according to a simple, modulo-based hash function. For example, let T be a node in the network and n be a level in the hierarchy. Let $\mathbf{ID}$ and S be the functions that respectively return the identity of a node and the set of nodes sharing the same region at the given level. The hash function can then be expressed as follows:

$$H(T, n) = \mathbf{ID}(T) \bmod \|S(T, n)\|$$

### 3.3. Mobility of Location Server

Due to the mobility of location servers, the location server of a node may no longer be its location server when it moves away from its region. This could be solved by: 1) the location server is responsible to find a new location server for the node in that region according to the above selection algorithm, and move the location information of the node to the new location server; 2) the location server just discards the location information of the node. Each time the node changes location, it may check whether the original location server is still in the region. In this case, it updates the location server with its new location information. Otherwise, it chooses a new server using the location server selection algorithm as described above.

### 3.4. Location Update

When a node leaves a region, it updates its local location server with its most recent location information (location and velocity).

### 3.5. Location Request

In order to send a message, a sender node needs to know the position of the destination node. The sender node contacts its local location server and queries the location of the destination node. If the real location of the destination node is known, it is returned to the sender node. Otherwise, in conjunction with a predicted location computed according to the prediction algorithm as described in the next section, the request is forwarded to the location server in the next upper level which broadcasts the request to the location servers on the next lower level. Using this mechanism, the request is forwarded from location servers to location servers until the real location of the destination node is found, or the highest-level location server has been reached. In this case, a predicted location is returned traversing the tree. To avoid the problem of expired predicted information, the predicted location is associated a timestamp and the most recent one is chosen.

### 3.6. Prediction Method

Two predictive methods have been used to study the performance of PHLS. These two methods are very similar and only differ in the way they are taking the velocity of nodes into account. Therefore, in the following, we will refer to the variants of PHLS as PHLS1 and PHLS2.

### 3.6.1. Predictive Velocity-based method (used in PHLS1)

The first method determins the possible location of a node using both the last known location and the velocity (including the direction) of this node, considering the probability for the node to change direction is quite low. Let l be the location of the node, v be its velocity and t be the time when l and s have been measured or computed. "rec" and "now" indices are used to indicate a recorded value and the present respectively. The predicted location of the node $l_{now}$ can then be expressed as follows:

$$l_{now} = l_{rec} + v_{rec} \times (t_{now} - t_{rec}) \qquad (1)$$

### 3.6.2. Moving Average Velocity Prediction method (used in PHLS2)

This second method is based on the formula presented in [23] and [24]. It uses an equation similar to (1) to determine the possible location of the node; however, the last recorded velocity is not taken into account directly but smoothed using previous velocities into an average velocity. Let $\bar{v}$ be the average velocity of the node and $\alpha$ be a filter gain constant used to compute the relative weights of the node's previous average velocity and the last recorded velocity. The new average velocity can be expressed as follows:

$$\bar{v}_{new} = \alpha \ \bar{v}_{old} + (1 - \alpha) \ v_{rec} \qquad (2)$$

which leads to the following expression for the computation of the predicted location for the node:

$$l_{now} = l_{rec} + \bar{v}_{new} \times (t_{now} - t_{rec}) \qquad (3)$$

Note that 1) the value of $\alpha$ is normalized in the range from 0 to 1, ie. the closer to 0 the value of $\alpha$, the more important the last recorded velocity of the node and the closer to 1 the value of $\alpha$, the less important the last recorded velocity of the node and 2) the last $\bar{v}_{new}$ becomes the new $\bar{v}_{old}$ every time a new location has to be predicted for the node.

## 4. QUALITATIVE COMPARISON

Table 1 shows the location services that have been discussed above. Type indicates whether the network area is flat or recursively divided into a hierarchy of smaller and smaller grids. Some location services provide Localized Information by maintaining a higher density or better quality of position information nearby the position of the node. This may be of importance when communications in an ad-hoc network are mainly local.

The Robustness of a location service is low, medium or high and depends on the kind of failure taken into account to return that a node is unaccessible, typically a single node, a small subset of all nodes or all nodes respectively. The implementation Complexity describes the inherent complexity of the location service and the level of difficulty to implement and test it. When the location service uses location servers, the Location Server Identification (LSI) specifies whether these servers are identified by their node id (ID-based LSI) or by their actual position (position-based LSI). The Area Division refers to the use of of a hierarchical structure, where the hierarchies are based on dividing the area into different regions and use sets of these regions to form regions of a higher level. Most location services using area divisions also use a hierarchical structure. Both Update Strategy and Request Strategy describe the methods used by the location service to find the location servers. Strategies can be flooding, geocast, unicast and treewalk. When using the treewalk, update and request packets are forwarded according to the treelike structure of the hierarchy, ie. following a branch from a leaf to the root. Finally, Table 1 shows that a priori, for the cost of a medium complexity (which complexity is inherent to the nature of hierarchical methods), PHLS1 and PHLS2 have good properties for the location of nodes. Essentially, they are robust to node failure and do not need to flood the network for both updates and requests.

Table 1: Comparison of studied location services

| Criterion | PLS | HLS | PHLS1,2 |
|---|---|---|---|
| Type | Flat | Hierarchical | Hierarchical |
| Localized Information | No | Yes | Yes |
| Robustness | High | Medium | High |
| Complexity | Low | Medium | Medium |
| LSI | - | Position-based | Position-based |
| Area Division | No | Yes | Yes |
| Update Strategy | Flooding | Geocast | Unicast |
| Request Strategy | Flooding | Treewalk | Treewalk |

## 5. MATHEMATICAL ANALYSIS

The framework as presented in [25] has been used to analyze the scalability of PHLS. The focus is set on how well the protocol scales when either or both the number of nodes in the network and the moving speed of nodes increase. Three metrics have been used: location maintenance cost, location query cost and storage cost.

Both location maintenance cost and location query cost are evaluated based on the forwarding load, i.e. the number of hops a packet needs to traverse during each operation (update/query). This way, packets travelling far away have a higher cost than packets sent to a nearby destination. Moreover, all three metrics are defined in terms of individual node. Since nodes are supposed to be symmetric in MANET, the expected values for the metrics are the same for all nodes in the network. Also, the following two assumptions are made:

· For a fixed node density, the area of the network grows linearly with the number of nodes.

· A data traffic pattern is the probability distribution of traffic intensities between any pair of nodes in the network. The uniform model pattern is used where the probability of initiating a packet between any pair of nodes is equal.

Table 2 summarizes the notations used for the analysis.

### 5.1. Definition of metrics

A location maintenance is performed each time a node crosses the boundary of the region where it is located to go to an adjacent one. When this occurs, the maintenance packet is forwarded to

Table 2: Notations

Parameters

| | |
|---|---|
| A | Area of the network |
| H | Number of hierarchical level |
| N | Number of nodes |
| R | Side length of a level-0 square |
| v | Node speed |

Variables

| | |
|---|---|
| z | Average progress of each forwarding hop |
| d | Average distance travelled by a node |
| $d_i$ | Distance travelled by a packet at level i |
| $p_i$ | Level-i square boundary crossing rate |
| c | Random distance within/between squares |
| $n_i$ | Number of forwarding hops for a packet at level i |
| $P_i$ | Probability of querying nodes in level-i square |

Metrics

| | |
|---|---|
| $C_m$ | Location maintenance cost |
| $C_q$ | Location query cost |
| $C_s$ | Storage cost |

the destination. As a result, the location maintenance cost can be expressed as follows:

$$E(C_m) = \sum_{i=0}^{H} \rho_i \, E(n_i) \qquad (4)$$

A location query is performed every time the destination node is not located in the same region as the sender node. When this occurs, the query packet is sent to higher-level regions. The location query cost can be expressed as follows:

$$E(C_q) = \sum_{i=0}^{H} P_i \, E(n_i) \qquad (5)$$

The storage cost is defined as the number of location records a node need to store as a location server. If the distribution of location servers is uniformly distributed among nodes, each node is the location server of exactly one other node for each level of the hierarchy. Therefore, the storage cost can be expressed as follows:

$$E(C_s) = H + 1 \qquad (6)$$

## 5.2. Evaluation of variables

As shown in (4), the location maintenance cost depends on both the boundary crossing rate at the different levels and the expected number of forwarding hops for an update packet.

The boundary crossing rate is different from one level to another. Typically, considering the hierarchical organization of the regions in which each region at a given level covers exactly four regions of the underlying level, the boundary crossing rate at a given level is four times less than the one of the underlying level, that is:

$$\rho_{i+1} = \frac{\rho_i}{4} \quad \text{or} \quad \rho_i = 4^{-i}\rho_0 \quad \forall i \quad 1 \leq i \leq H \tag{7}$$

At the lowest level, the boundary crossing rate depends on the mean distance travelled by packets in regions. This distance is extremely difficult to estimate for square regions due to the inherent structure of squares that only includes few axial symetries. However, it is quite straightforward for a circle which inherent structure includes a central symetry. Therefore, as this study focusses on the order of metrics and not on their strict evaluation and as considering a region as being a circle instead of being a square has no impact on the order of the metrics, a circle is used for the evaluation of the mean distance travelled by packets in regions instead of a square. In a circle, d depends on the diameter of the circle and the angle of the packet path with regards to the tangent of the circle at the entry point:

$$d = \frac{\int_0^{\pi/2} R\cos\theta\, d\theta}{\pi/2} = \frac{2R}{\pi} \tag{8}$$

At level 0, the boundary crossing rate can be expressed as the ratio of the node's speed by the mean distance travelled by the node in the cell, ie. $p_0 = v/d$. The boundary crossing rate at level i is then given by:

$$\rho_i = \frac{1}{4^i} \times \frac{\pi v}{2R} \quad \forall i \quad 0 \leq i \leq H \tag{9}$$

The average number of forwarding hops for a packet at level i depends on the average distance travelled by the packet at this level and the average progress performed by the packet at each hop. This progress depends on physical characteristics, typically the transmission range of antennas and the node density. The transmission range is a constant for our model and so is the node density as defined by the ratio between the network area and the number of nodes. As a result, z is a constant and the average number of forwarding hops for a packet at level i can be expressed as follows:

$$E(n_i) = \frac{E(d_i)}{z} \quad \forall i \quad 0 \leq i \leq H \tag{10}$$

The average value for the distance travelled by a packet at level i is the mean distance between any two random points in a level i square. At any level i, $E(d_i)$ can be expressed by:

$$E(d_i) = 4^i Rc \quad \forall i \quad 0 \leq i \leq H \tag{11}$$

where c is a constant defined by:

$$c = \int_0^1\int_0^1\int_0^1\int_0^1 \sqrt{(x_1-x_2)^2+(y_1-y_2)^2}\, dx_1 dx_2 dy_1 dy_2 \tag{12}$$

The location query cost as defined in (5) depends on $E(n_i)$ too and also on the probability that a query is satisfied at level i. Considering that any regions are divided into four sub-regions, the probability for a query to be satisfied at level i is straightforwardly provided by:

$$P_i = \begin{cases} \dfrac{3}{4^{H-i}} & \forall i \quad 1 \leq i \leq H \\ \dfrac{1}{4^H} & \text{if } i = 0 \end{cases} \tag{13}$$

## 5.3. Metrics evaluation

Both location maintenance cost and location query cost can be evaluated from equations (9) to (12) and from equations (10) to (13) respectively:

$$E(C_m) = \frac{v \pi c H}{2z} \qquad E(C_q) \simeq \frac{3 \times 4^H c R}{z} \qquad (14)$$

As c, Z and R are all constants, the order of $E(C_m)$ depends on the order of both H and v while the order of $E(C_q)$ depends on the order of H but is independent of v. This is also the case for $E(C_s)$. Thus:

$$E(C_m) = O(vH) \quad E(C_q) = O(4^H) \quad E(C_s) = O(H) \qquad (15)$$

If the side length of level-0 squares is kept constant, the number of hierarchical level is proportional to the side length of the network area. Then, if the density of nodes is kept constant, the network area is proportional to the number of nodes in the network. As a result:

$$\left. \begin{array}{l} H \propto \log \sqrt{A} \\ A \propto N \end{array} \right\} \rightarrow H \propto \log \sqrt{N} \qquad (16)$$

From (16), equations in (15) can then be expressed as a function of N:

$$\begin{array}{l} E(C_m) = O(v \log N) \\ E(C_q) = O(\sqrt{N}) \\ E(C_s) = O(\log N) \end{array} \qquad (17)$$

By definition, a system is scalable according to a metric when the order of the metric grows smaller than the order of the parameters. As all the metrics have an order smaller than N and v, it is possible to state that our proposal is scalable when the number of nodes and/or the node's velocity vary.

## 6. PERFORMANCE EVALUATION

We implemented PHLS1 and PHLS2, and compared them to the HLS protocol. Our simulations were conducted without any data traffic on the network, resulting in queries being sent out instead of data, and the greedy forwarding routing protocl was used.

Table 3 lists the simulation parameters that are used. The simulation area is 1000 m × 1000 m. The transmission range is set to 250 m. The mobility model is the modified random direction mobility model [26] where a node arbitrary chooses its speed, time and direction and moves into that direction until the chosen time period has expired. The node pauses for awhile and then other values for these parameters are selected. If the node hits the border of the simulation area, it bounces back according to the physical laws of reflection. The pause time is set to 10 seconds.

## 6.1. Impact of nodes' mobility

The first scenario aims at studying the impact of nodes' mobility on various metrics (query success rate, prediction accuracy, etc.) for the evaluated protocols. The number of nodes is fixed to 300. The maximum velocity for nodes is taken in the range from 10 m.s$^{-1}$ to 50 m.s$^{-1}$, ie. nodes select their moving speed in the range 0 to 50 m.s$^{-1}$ for the latter case and pause for a maximum of 10 s before changing direction.

Table 3: Simulations parameters

| Input Parameters | |
|---|---|
| Simulation area | 1000 m × 1000 m |
| Transmission range | 250 m |
| Number of nodes | 100, 200, 300, 400 |
| Mobility model | Modified Random Direction |
| Max speed | 10, 20, 30, 40, 50 m/s |
| Pause time | 10 s |
| Number of requests per run | 1200 |

| Network Simulator | |
|---|---|
| Simulator | NS-2 v.2.29 |
| Medium Access Protocol | IEEE 802.11 |
| Link bandwidth | 2 Mbps |
| Simulation duration | 300 s |
| Number of runs | 5 |

### 6.1.1. Query Success Rate

This is the percentage of queries that are successfully resolved by the location servers. Fig. 1 shows that the query success rate decreases as the speed increases. PHLS2 provides the highest percentage of answered location requests and PHLS1 outperforms HLS.

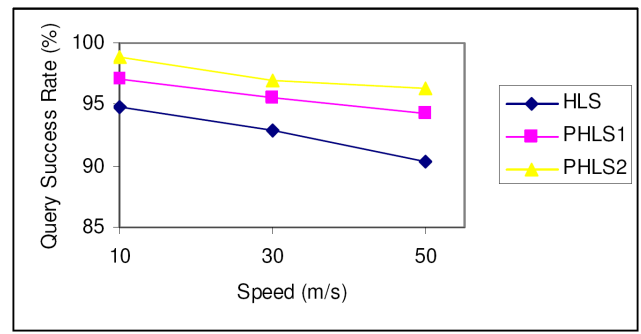

Figure 1: Query success rate vs. speed.

The query success rate decreases for both PHLS1 and PHLS2. This is related to the fact that nodes change their direction in smaller time intervals and our implemented velocity prediction schemes do not consider the direction of node movement. Regarding HLS, the query success rate decreases at higher speed. This is due to the number of query failures that arise when a node moves far away from its previous location leading to an invalid stored location in location servers.

### 6.1.2. Prediction accuracy (Average Location Error)

This is the absolute difference between the actual and the reported location of the node as stored in HLS or predicted in PHLS, and is measured in meters. As expected, Fig. 2 shows that the error increases as the speed increases for all location services.

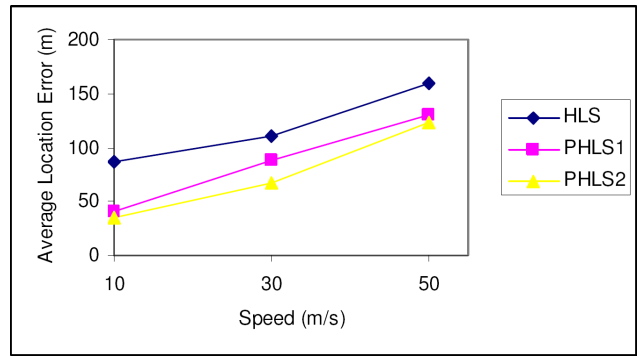

Figure 2: Average Location Error vs. speed.

Moreover, Fig. 2 also shows that PHLS2 predicts a more accurate location than PHLS1, and both outperform HLS. In fact, location errors in HLS occur when stale locations are stored in location servers, especially in high mobility environment. The use of prediction significantly limits these errors in PHLS1 and PHLS2.

### 6.1.3. Bandwidth Consumption

This is the traffic overhead involved by the use of location servers. It is measured as the amount of data transmitted per period of time and per node, ie. in $byte.s^{-1}.node^{-1}$. As shown in Fig. 3, the bandwidth consumption increases as the speed increases for all location services. This is due to the fact that as the speed increases, the probability for a node to change cell and/or region is getting higher.

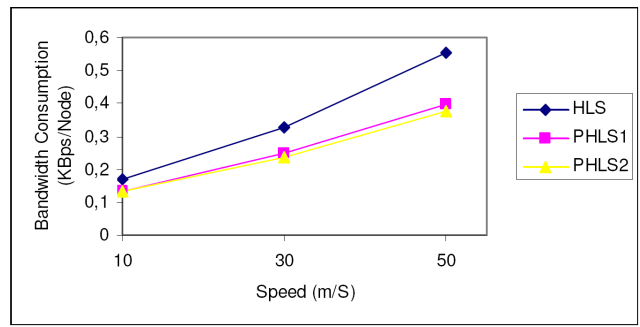

Figure 3: Bandwidth Consumption vs. speed.

Moreover, Fig. 3 also shows that the bandwidth consumption for PHLS1 and PHLS2 is lower than the one for HLS and that the higher the velocity of nodes, the bigger the difference between HLS on one hand and PHLS1 and PHLS2 on the other hand. The difference between HLS and PHLS is mainly due to the prediction. When HLS needs to send lots of messages to find a node that changed cell or region, PHLS limits the number of messages with the prediction of the node's location.

## 6.2. Impact of the node density

The second scenario aims at studying the impact of varying the node density on the performance of both HLS and PHLS protocols. Typically, it would be interesting to determine the minimum number of nodes in the area to make the use of HLS and PHLS a success. The speed is fixed at $10\,\mathrm{m.s}^{-1}$. The number of nodes varies from 100 to 400.

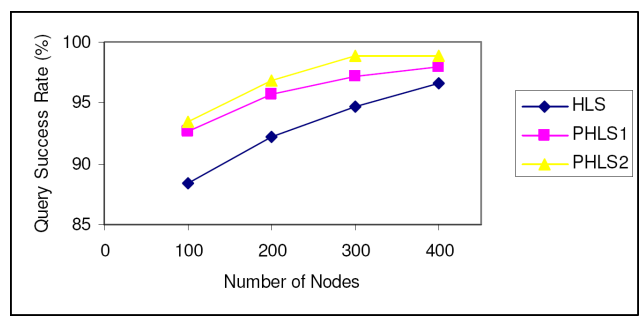

Figure 4: Query Success Rate vs. Number of Nodes.

Fig. 4 shows that the query success rate for the evaluated protocols increases as the node density increases. The reasons for the query failure at a low density is due to the route failure with the simple greedy forwarding routing. Moreover, HLS achieves a lower success rate than PHLS1 and PHLS2 due to empty cells that cause lost of updates which leads to invalid location information in the location servers.

## 7. CONCLUSION AND FUTURE WORKS

This paper presented the Predictive Hierarchical Location Service (PHLS). PHLS uses a hierarchy of cells and regions to achieve scalability and predicts the requested location when the exact location is unknown by using a previously stored location information (both location and velocity). By combining the hierarchy structure together with the prediction, flooding is avoided (despite in PLS) and the performance are improved.

Two prediction schemes have been used to evaluate our new location service. Comparisons and analysis of the performance of PHLS with regards to those of HLS showed that PHLS outperforms HLS for all studied metrics whether the speed and/or the number of nodes varies. Results are different for the two prediction methods and, even though results are better with the Moving Average Velocity Prediction method rather than the Predictive Velocity-Based method, both are providing better results that HLS. Moreover, the mathematical analysis of our proposal showed its good scalability when both the number of nodes in the network and the velocity of the nodes vary.

As of now, both presented prediction schemes do not take into account direction changes. As a future work, we intend to enhance the predictive schemes by taking into account the direction changes of node movement. Our on-going work also consists in studying PHLS with deep simulation analysis under different mobility models.

Authors

Ebtisam Amar has been a PhD student at University Pierre and Marie Curie, Paris, France, since 2006. Her research interests include wireless and mobile networks and especially node localization. Ebtisam holds a MSc in Computer Science from Institut Télécom – Télécom SudParis, Évry, France.

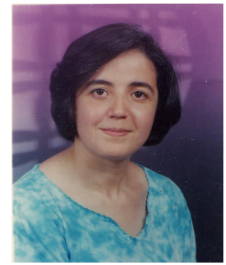

Selma Boumerdassi has been an associate professor at the Conservatoire National des Arts et Métiers, Paris, France, since 2000. Her research interests include wireless and mobile networks especially for routing and security, and RFID systems. She received a MSc in Computer Engineering in 1993 from the Institut National d'Informatique, Algeria, a MSc in Computer Science in 1995 and a PhD in Mobile Networks in 1999 from the Université de Versailles-Saint-Quentin-en-Yvelines, France, where she also served as an assistant professor.

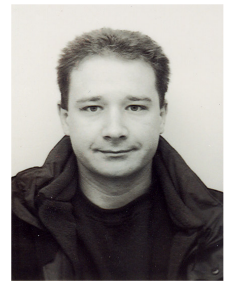

Éric Renault has been an associate professor at the Institut Télécom – Télécom SudParis, Évry, France, since late 2001. His research interests include cluster and grid computing, high-performance messaging together with low-cost security and the development of the Future Internet. He has been involved in several national (SecResCap, System@tic Carriocas, PerCo and AMOS) and European projects (FP7 IP 4WARD, FP7 STREP Mobesens and FP7 NoE Euro-NF). Éric received a MSc in Computer Engineering and a MSc in Computer Science in 1995 and a PhD in Parallel Computing in 2000 from the University of Versailles Saint-Quentin-en-Yvelines, France, where he also served as an assistant professor. In 2001, he was a research associate at Dartmouth College, NH. Éric is the author of more than 50 papers published in peer-reviewed journals and conferences.